\documentclass[journal,twoside,web]{ieeecolor}
\usepackage{generic}
\usepackage{cite}
\usepackage{amsmath,amssymb,amsfonts}
\usepackage{algorithmic}
\usepackage{booktabs} 
\usepackage{pifont}
\usepackage{multirow}
\usepackage{url}
\usepackage{pifont}
\usepackage{array}  
\usepackage{graphicx}
\usepackage{threeparttable} 
\usepackage{algorithm,algorithmic}
\usepackage{hyperref}
\newcommand{\xmark}{\ding{55}}
\usepackage{textcomp}
\def\BibTeX{{\rm B\kern-.05em{\sc i\kern-.025em b}\kern-.08em
    T\kern-.1667em\lower.7ex\hbox{E}\kern-.125emX}}
\begin{document}

\title{Multimodal Latent Fusion of ECG Leads for Early Assessment of Pulmonary Hypertension}

\author{Mohammod N. I. Suvon, Shuo Zhou, Prasun C Tripathi, Wenrui Fan, Samer Alabed, Bishesh Khanal, Venet Osmani, Andrew J Swift, Chen Chen, and Haiping Lu, \IEEEmembership{Senior Member, IEEE}
\thanks{Mohammod N. I. Suvon, Shuo Zhou, Prasun C Tripathi, Wenrui Fan, Chen Chen, and Haiping Lu are with School of Computer Science, University of Sheffield, S1 4DP, Sheffield, U.K. (Corresponding Author: Haiping Lu, E-mail: h.lu@sheffield.ac.uk).}
\thanks{Mohammod N. I. Suvon, Shuo Zhou, Prasun C Tripathi, Wenrui Fan, and Haiping Lu are also with Centre for Machine Intelligence, University of Sheffield, S1 3JD, Sheffield, U.K.}
\thanks{Prasun C Tripathi is also with Department of Electrical \& Computer Science Engineering, IITRAM Ahmedabad, Gujarat-380 026, India.}
\thanks{Venet Osmani is with Digital Environment Research Institute, Queen Mary University of London, E1 1HH, London, U.K.}
\thanks{Bishesh Khanal is with Nepal Applied Mathematics and Informatics Institute for research (NAAMII), Nepal, Jwagal, Lalitpur-44700, Nepal.}
\thanks{Chen Chen is also with Department of Computing, Imperial College London, SW7 2AZ, London, U.K.}
\thanks{Samer Alabed and Andrew J. Swift are with School of Medicine and Population Health, University of Sheffield, S10 2TN Sheffield, Department of Clinical Radiology, Sheffield Teaching Hospitals, S10 2JF, Sheffield, U.K, and National Institute for Health and Care Research (NIHR), Sheffield Biomedical Research Centre, S10 2JF, Sheffield, U.K.}
}

\maketitle

\begin{abstract}
Recent advancements in early assessment of pulmonary hypertension (PH) primarily focus on applying machine learning methods to centralized diagnostic modalities, such as 12-lead electrocardiogram (12L-ECG). Despite their potential, these approaches fall short in decentralized clinical settings, e.g., point-of-care and general practice, where handheld 6-lead ECG (6L-ECG) can offer an alternative but is limited by the scarcity of labeled data for developing reliable models. To address this, we propose a lead-specific electrocardiogram multimodal variational autoencoder (\textsc{LS-EMVAE}), which incorporates a hierarchical modality expert (HiME) fusion mechanism and a latent representation alignment loss. HiME combines mixture-of-experts and product-of-experts to enable flexible, adaptive latent fusion, while the alignment loss improves coherence among lead-specific and shared representations. To alleviate data scarcity and enhance representation learning, we adopt a transfer learning strategy: the model is first pre-trained on a large unlabeled 12L-ECG dataset and then fine-tuned on smaller task-specific labeled 6L-ECG datasets. We validate \textsc{LS-EMVAE} across two retrospective cohorts in a 6L-ECG setting: $892$ subjects from the ASPIRE registry for (1) PH detection and (2) phenotyping pre-/post-capillary PH, and $16,416$ subjects from UK Biobank for (3) predicting elevated pulmonary atrial wedge pressure, where it consistently outperforms unimodal and multimodal baseline methods and demonstrates strong generalizability and interpretability. The code is available at \textcolor{blue}{\url{https://github.com/Shef-AIRE/LS-EMVAE}}.
\end{abstract}
\section{Introduction}
\label{sec:introduction}
\IEEEPARstart{P}{ulmonary} hypertension (PH) is a complex and heterogeneous condition that affects approximately 75 million people worldwide~\cite{maarman2020pulmonary}, including up to 10\% of those over the age of 65, who face a poor prognosis with a low 5-year survival rate~\cite{hoeper2016global}. In clinical practice, PH is diagnosed and phenotyped using cardiac hemodynamics obtained through right heart catheterization (RHC) in centralized clinical settings, such as hospitals, specialty clinics, diagnostic centres, and clinical laboratories. Two key cardiac hemodynamics that serve as surrogate markers in PH assessment are mean pulmonary arterial pressure (mPAP) for detecting PH and pulmonary artery wedge pressure (PAWP) for phenotyping pre-/post-capillary PH~\cite{humbert20222022}. Although RHC is the gold standard, it is invasive, complex, and expensive, which makes it unsuitable for large-scale PH screening. For example, in China, fewer than 7\% of hospitals are equipped to perform RHC, limiting access to definitive PH diagnosis~\cite{ma2021medical}.

In recent years, many machine learning (ML)-based methods have been developed to predict cardiac hemodynamics using high-cost non-invasive imaging modalities such as cardiac magnetic resonance imaging (CMRI)\cite{tripathi2023tensor, tripathi2024interpretable} and echocardiography\cite{zhao2025development, traversi2001doppler,yang2025ai}, as well as low-cost modalities like the 12-lead electrocardiogram (12L-ECG)\cite{aras2023electrocardiogram, schlesinger2022deep, dubrock2024electrocardiogram} and chest X-rays (CXR)\cite{kusunose2020deep}, in centralized clinical settings. For instance, Aras et al.~\cite{aras2023electrocardiogram} showed that an automated ML approach using 12L-ECG alone could detect PH and its phenotypes. Suvon et al.~\cite{suvon2024multimodal} showed that CXR and 12L-ECG can be used jointly to detect cardiac hemodynamic instability, where 12L-ECG outperformed CXR in unimodal comparison, highlighting its effectiveness in capturing hemodynamic markers relevant to PH.

Despite these developments, PH diagnosis remains challenging due to its complex causes and subtle early symptoms. This often results in delays of 2.5 to almost 4 years from symptom onset to confirmed diagnosis~\cite{khou2020diagnostic, didden2023time}. Such delays can be prolonged or even cause misdiagnosis in decentralized clinical settings, such as point-of-care, general practice, and home-based care, where experienced clinicians and advanced diagnostic tools are scarce and clinicians solely rely on symptom-based evaluation. This emphasizes the need for portable diagnostic tools for PH assessment, such as handheld 6-lead ECG (6L-ECG)~\cite{kardiamobile6L}, which records the six limb leads (I, II, III, aVR, aVL, aVF).

Several studies~\cite{lim2025artificial, dwivedi2023machine, bacevicius2023six} have demonstrated the potential of 6L-ECG for diagnosing various cardiac conditions using ML models, but its value for PH remains largely unexplored. Moreover, the scarcity of labeled 6L-ECG data remains a major challenge, limiting the development of robust supervised models, which require lots of training data. To address this, recent studies~\cite{na2024guiding, zhang2022maefe} have explored semi-supervised pre-training approaches that leverage large-scale unlabeled data. For example, Na et al.~\cite{na2024guiding} introduced a model pre-trained on unlabeled 12L-ECGs and demonstrated its adaptability to labeled 6L-ECGs for arrhythmia classification. McKeen et al.~\cite{mckeen2024ecg} proposed a foundation model ECG-FM pre-trained on 1.5 million unlabeled ECGs and fine-tuned on various downstream tasks such as left ventricular ejection fraction prediction and atrial fibrillation.

Most aforementioned methods consider multi-lead ECG as a single modality and primarily use unimodal approaches for lead fusion. In the literature, ML-based methods for integrating ECG leads can be broadly classified into two types. The first type is simple input-level lead fusion~\cite{hong2020holmes,strodthoff2020deep,vaswani2017attention}, where all ECG leads are concatenated or stacked into a joint sequence before encoding. The second type is encoder-level lead fusion~\cite{na2024guiding,mckeen2024ecg,yue2022ts2vec,das2024decoder}, where each lead is first processed separately and then combined within the encoder to learn joint representations. This approach is widely used in Transformer-based models with self-attention mechanisms. For instance, ST-MEM~\cite{na2024guiding} applies positional and spatial encoding to each lead and performs integration through attention layers in the later stages of the encoder. In contrast, TS2Vec~\cite{yue2022ts2vec} integrates the leads at the earlier stages of the encoder to jointly capture spatial and temporal patterns.

However, among existing ECG lead fusion methods, only a few incorporate lead-specific information and demonstrate the varying diagnostic importance of different leads. This is especially important in PH, where some leads carry more predictive value than others. For instance, in PH detection using 6L-ECG, Lead-II is particularly more informative as it provides a clear view of atrial depolarization, enabling the detection of right atrial enlargement and pressure changes~\cite{henkens2008pulmonary}. Prior multimodal latent fusion techniques~\cite{wu2018multimodal,palumbo2023mmvae+,sutter2021generalized} have shown success in combining different modalities, but their application to ECG lead fusion remains largely unexplored.

To address all the above-mentioned challenges (i.e., the unmet need for 6L-ECG in early PH assessment, the scarcity of labeled 6L-ECG data, and the limitations of unimodal ECG lead fusion), we propose \textsc{LS-EMVAE}, a lead-specific electrocardiogram multimodal variational autoencoder model. \textsc{LS-EMVAE} is designed to pre-train on 12L-ECG data to learn transferable latent representations and then fine-tune on 6L-ECG data. Specifically, for pre-training, we first treat each ECG lead as a separate modality and use individual encoders to extract its latent distribution. Second, the encoded latent distributions are passed through a hierarchical modality expert (HiME) fusion module, which has two levels: a product-of-experts (PoE) that combines latent distributions from individual leads into a shared distribution, and a mixture-of-experts (MoE) that fuses the shared and lead-specific distributions into a joint representation. To further enhance this fusion, we introduce a latent representation alignment (LRA) loss that improves the coherence among lead-specific and shared representations. The joint representation is then decoded using a shared decoder to learn cross-lead representations.

Finally, during fine-tuning, the decoder is removed, the 6L-ECG encoders are retained, and a task-specific classifier is attached for three downstream tasks: PH detection, PH phenotyping, and elevated PAWP prediction. We conduct extensive experiments on two retrospective cohorts, ASPIRE registry~\cite{hurdman2012aspire} and UK Biobank~\cite{petersen2016uk}, using a 6L-ECG setting, comparing against seven state-of-the-art (SOTA) unimodal models~\cite{hong2020holmes, strodthoff2020deep, vaswani2017attention, van2023joint, mckeen2024ecg, na2024guiding, das2024decoder} and three multimodal latent fusion models~\cite{wu2018multimodal, palumbo2023mmvae+, sutter2021generalized}. The results demonstrate that \textsc{LS-EMVAE} consistently outperforms all competitive models and offers detailed interpretability via lead-level and wave-level integrated gradient attribution ratio (IGAR) analysis. Notably, while 12L-ECG achieves higher performance, 6L-ECG offers a cost-effective alternative, underscoring a balanced trade-off between performance and scalability for PH detection and phenotyping in decentralized clinical settings.


To summarize, this study differs from prior work in three main ways. First, to the best of our knowledge, we are the first to explore the trade-off between 12L-ECG and 6L-ECG in decentralized clinical settings, establishing 6L-ECG as a cost-effective alternative for PH detection and phenotyping. Second, the proposed LS-EMVAE integrates individual lead and shared information through multimodal pre-training and is further regularized with a latent representation alignment loss, which together improve coherence and information exchange among individual leads and shared representation. Third, we introduce IGAR to provide quantitative interpretability of leads and waveforms, showing clinically relevant insights that support the model’s predictions in PH assessment.

\begin{figure*}[!t]
  \centering
      \includegraphics[width=\textwidth, trim=0cm 0cm 1cm 0cm]{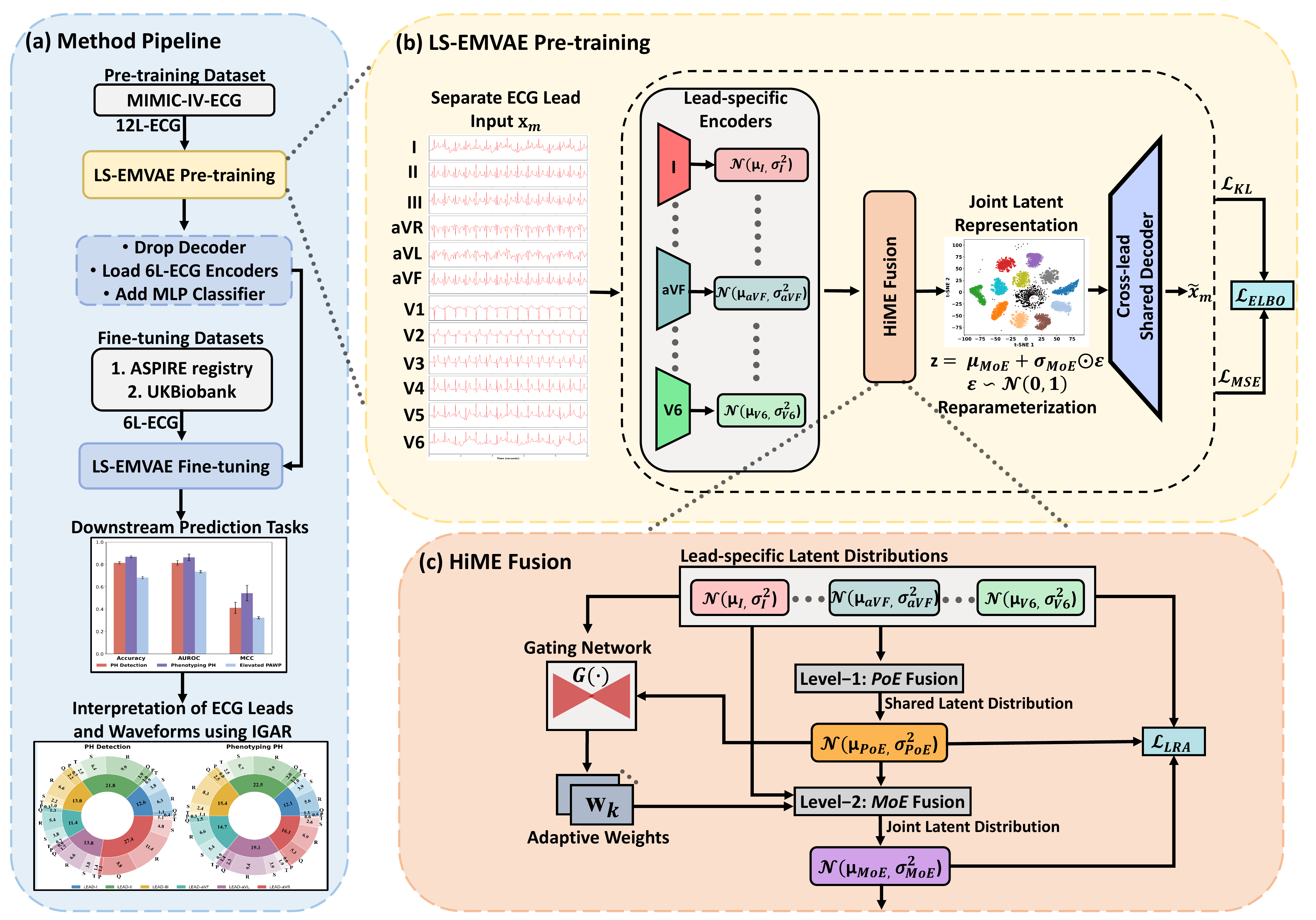}
  \caption{(a) Left: An overview of the proposed LS-EMVAE pipeline, which comprises pre-training, fine-tuning, and prediction on three downstream classification tasks: PH detection, phenotyping PH, and elevated PAWP prediction (see Section~\ref{sec:dataset} for task descriptions). Integrated Gradient Attribution Ratio (IGAR) is adopted for model interpretation. (b) Upper right: LS-EMVAE pre-training stage, where lead-specific encoders take individual ECG lead signals $x_m$ as input to generate separate latent distributions. These distributions are fused using a hierarchical modality expert (HiME) module to form a joint latent distribution. A latent variable is then sampled from this fused distribution using the reparameterization trick to obtain the joint latent representation, which is subsequently passed through a cross-lead shared decoder to reconstruct the ECG leads $\hat{x}_m$. Here, \( m \) denotes the individual ECG leads \(\{I, II, III, aVR, aVL, aVF, V_1, V_2, V_3, V_4, V_5, V_6\}\), and \( k \in \{m, \text{PoE}\} \) denotes all leads together with PoE. (c) The HiME fusion module performs fusion in two levels: at Level 1, PoE combines the individual lead distributions into a shared distribution, and at Level 2, MoE integrates the shared and lead-specific distributions into a joint latent distribution. The latent representation alignment (LRA) loss further enhances coherence among them.}
  \label{pipeline}
\end{figure*}

\section{Methods}
\label{sec:methods}
This section presents our proposed LS-EMVAE pipeline, describing its overall structure and components. As shown in Fig.~\ref{pipeline}(a), the method pipeline is organized into two stages. The first stage involves pre-training on 12L-ECG data to learn generalizable joint latent representations, as illustrated in Fig.~\ref{pipeline}(b). In the second stage, the pretrained model is fine-tuned on 6L-ECG data for task-specific classification. We also perform quantitative interpretability analysis using IGAR to evaluate the contribution of each ECG lead and waveform to the model’s predictions.

\subsection{LS-EMVAE Architecture}
LS-EMVAE consists of three main components: Encoder, HiME-Fusion, and Decoder, as shown in Fig.~\ref{pipeline}(b).
\subsubsection{Encoder Design}
In the encoder module, each lead of the 12-lead ECG is treated as a separate modality and processed independently by a dedicated convolutional neural network (CNN), resulting in 12 encoders. Each encoder processes a lead signal $\mathbf{x}_m \in \mathbb{R}^{L}$, where $L$ denotes the signal length and $m \in \{\text{I}, \text{II}, \text{III}, \text{aVR}, \text{aVL}, \text{aVF}, \text{V1}, \text{V2}, \text{V3}, \text{V4}, \text{V5}, \text{V6}\}$. It consists of three 1D convolutional layers with kernel size $1 \times 3$, stride $2$, and padding $1$, each followed by a ReLU activation, with channel depths of $16$, $32$, and $64$, respectively, producing a feature map. This feature map is flattened and passed through two fully connected layers to produce the latent expert mean $\boldsymbol{\mu}_m$ and variance $\boldsymbol{\sigma}_m^2$, which parameterise a latent distribution $\mathcal{N}(\boldsymbol{\mu}_m, \boldsymbol{\sigma}_m^2)$.

\subsubsection{HiME-Fusion Module}
The HiME-Fusion module is a two-level hierarchical fusion method that consists of three main components, as shown in Fig.~\ref{pipeline}(c). The first level fusion and initial component is the product of experts (PoE)~\cite{wu2018multimodal}, which takes the individual lead latent distributions \( \mathcal{N}(\boldsymbol{\mu}_m, \boldsymbol{\sigma}_m^2) \) as input and combines them into a shared latent distribution \( \mathcal{N}(\boldsymbol{\mu}_{\text{PoE}}, \boldsymbol{\sigma}_{\text{PoE}}^2) \). The experts of the PoE latent distribution are computed as follows:
\begin{equation}
\begin{aligned}
\boldsymbol{\mu}_{\text{PoE}} &= 
\cfrac{\sum\limits_{m} \boldsymbol{\mu}_m / \boldsymbol{\sigma}_m^2}
      {\sum\limits_{m} 1 / \boldsymbol{\sigma}_m^2}, \\
\boldsymbol{\sigma}_{\text{PoE}}^2 &= 
\left( \sum\limits_{m} \frac{1}{\boldsymbol{\sigma}_m^2} \right)^{-1}.
\end{aligned}
\end{equation}

The second component of this module is a gating network, denoted as \( G (\cdot) \), which takes the latent experts \( \boldsymbol{\mu}_k \), where \( k \in \{m, \text{PoE}\} \), representing both the individual lead and shared latent experts. The gating network outputs a set of adaptive weights \(w_k\) that determine the contribution of each lead and the shared latent distribution to the joint latent distribution, thereby adaptively suppressing weak or noisy experts. The adaptive weights are obtained by applying a softmax function to the gating network outputs:
\begin{equation}
w_k = \frac{\exp(G(\boldsymbol{\mu}_k))}{\sum_{k} \exp(G(\boldsymbol{\mu}_k))},
\end{equation}
where \( G(\cdot) \) is a lightweight multi-layer perceptron (MLP) network. The softmax function ensures that the weights are non-negative and sum to 1 across all experts.

Finally, the second-level fusion, which is the third component of HiME-Fusion, uses a mixture of experts (MoE)~\cite{palumbo2023mmvae+} to fuse the individual lead \( \mathcal{N}(\boldsymbol{\mu}_m, \boldsymbol{\sigma}_m^2) \) and the shared \( \mathcal{N}(\boldsymbol{\mu}_{\text{PoE}}, \boldsymbol{\sigma}_{\text{PoE}}^2) \) latent distribution, weighted by the adaptive weights \(w_k\) learned for each distribution. This fusion results in a joint latent distribution \( \mathcal{N}(\boldsymbol{\mu}_{\text{MoE}}, \boldsymbol{\sigma}_{\text{MoE}}^2) \). The experts of the MoE latent distribution are computed as:
\begin{equation}
\begin{aligned}
\boldsymbol{\mu}_{\text{MoE}} &= \sum_{k} \boldsymbol{w}_k \cdot \boldsymbol{\mu}_k, \\
\boldsymbol{\sigma}^2_{\text{MoE}} &= \sum_{k} \boldsymbol{w}_k \cdot (\boldsymbol{\sigma}_k^2 + \boldsymbol{\mu}_k^2) - \boldsymbol{\mu}_{\text{MoE}}^2.
\end{aligned}
\end{equation}
The joint latent distribution \( \mathcal{N}(\boldsymbol{\mu}_{\text{MoE}}, \boldsymbol{\sigma}_{\text{MoE}}^2) \) is then reparameterized to sample the joint latent representation:
\begin{equation}
\mathbf{z} = \boldsymbol{\mu}_{\text{MoE}} + \boldsymbol{\sigma}_{\text{MoE}} \cdot \boldsymbol{\epsilon}.
\end{equation}
Here, \( \boldsymbol{\epsilon} \sim \mathcal{N}(0, \mathbf{I}) \) is a standard gaussian prior~\cite{bu2018estimation} used in the reparameterization trick~\cite{kingma2013auto} to obtain regularized and uncertainty-aware latent representations, where \( \mathbf{I} \) denotes the identity matrix.

\subsubsection{Decoder Design}
Our decoder module employs a cross-lead shared decoder network to reconstruct each lead of the 12L-ECG from the joint latent representation \( \mathbf{z} \). The decoder consists of a fully connected layer that maps \( \mathbf{z} \) to the reconstructed output $\tilde{\mathbf{x}}_m$, followed by three 1D transposed convolutional layers with kernel size 4, stride 2, and padding 1. These layers progressively restore the temporal structure of the corresponding ECG lead signal. The final layer applies an identity activation~\cite{godfrey2019evaluation}, ensuring that the reconstructed waveform maintains the correct amplitude scale. Sharing the decoder across leads encourages cross-lead generalization, ensures consistent reconstruction when leads are missing, and enables efficient fine-tuning by learning lead-invariant features that transfer well to different lead combinations and reduced-lead settings (e.g., 6L-ECG).

\subsection{LS-EMVAE Pre-training}
The LS-EMVAE model is pre-trained in an unsupervised manner on large-scale, unlabelled 12-lead ECG data. The overall pre-training objective combines two terms: (a) the Evidence Lower Bound (ELBO) loss, and (b) a latent representation alignment loss.

\subsubsection{Evidence Lower Bound (ELBO)}
During pre-training of LS-EMVAE, we optimize the Evidence Lower Bound (ELBO) objective, which consists of a reconstruction loss and a Kullback–Leibler (KL) divergence regularization term. The KL divergence term helps prevent variance collapse, a common issue in variational autoencoders, by encouraging the latent distribution to remain close to the gaussian prior~\cite{bu2018estimation}, thereby ensuring meaningful and disentangled representations~\cite{alemi2018fixing}. The ELBO loss is formulated as:
\begin{equation}
\label{ELBO}
\begin{aligned}
\mathcal{L}_{\text{MSE}} &= \sum_{m} \lambda_m \cdot \left\| \tilde{\mathbf{x}}_m - \mathbf{x}_m \right\|^2, \\
\mathcal{L}_{\text{KL}} &= \frac{1}{2} \left( \boldsymbol{\mu}_{\text{MoE}}^2 + \boldsymbol{\sigma}_{\text{MoE}}^2 - \ln(\boldsymbol{\sigma}_{\text{MoE}}^2) - 1 \right), \\
\mathcal{L}_{\text{ELBO}} &= \mathcal{L}_{\text{MSE}} + \beta \cdot \mathcal{L}_{\text{KL}},
\end{aligned}
\end{equation}
where the reconstruction loss \( \mathcal{L}_{\text{MSE}} \) is computed per ECG lead using mean squared error (MSE), weighted by \(\lambda_m\)~\cite{lawry2023multi}. The KL divergence \( \mathcal{L}_{\text{KL}} \) is scaled by a hyperparameter \(\beta\)~\cite{higgins2016beta}, which is gradually increased from 0 to 1 during training to smoothly enforce the prior constraint and improve latent representation regularization.

\subsubsection{Latent Representation Alignment}
We introduce a latent representation alignment loss (\(\mathcal{L}_{\text{LRA}}\)) that applies a simple yet effective MSE penalty to encourage each latent expert \( \boldsymbol{\mu}_k \) to remain consistent with its fused counterpart \( \boldsymbol{\mu}_{\text{MoE}} \) within the joint latent representation. The loss is defined as:
\begin{equation}
\label{LRA}
\mathcal{L}_{\text{LRA}} = \gamma \cdot \frac{1}{N} \sum_{k} \left\| \boldsymbol{\mu}_k - \boldsymbol{\mu}_{\text{MoE}} \right\|^2_2,
\end{equation}
where \(N\) is the total number of latent experts indexed by \(k\), and \(\gamma\) is a hyperparameter that controls the strength of the alignment.

The total pre-training objective combines Eq.~\ref{ELBO} and Eq.~\ref{LRA}:
\begin{equation}
\mathcal{L}_{\text{total}} = \mathcal{L}_{\text{ELBO}} + \mathcal{L}_{\text{LRA}},
\end{equation}
optimizing both individual and shared representations while maintaining alignment within the joint latent representation.

\subsection{LS-EMVAE Fine-tuning}
Fine-tuning the pretrained LS-EMVAE model is straightforward. For reduced-lead settings such as 6L-ECG, we remove the decoder, load the six limb lead encoders, and freeze them to retain the representations learned during pre-training. A two-layer MLP classifier is then attached to produce task-specific outputs, and binary cross-entropy loss is used for optimization. This design supports efficient transfer learning and enables generalization across different lead configurations.

\subsection{Integrated Gradient Attribution Ratio (IGAR)}
To enhance model interpretability, we introduce IGAR as a simple and scalable metric derived from Integrated Gradients (IG)~\cite{sundararajan2017axiomatic}. We compute IGAR for ECG signals at two levels: lead-level and wave-level.

For each ECG lead \( m \), we compute the IG attribution scores \( \boldsymbol{\alpha}_m \in \mathbb{R}^L \) with respect to the predicted class. The attribution scores are min–max normalized to yield \( \tilde{\boldsymbol{\alpha}}_m \). Salient time points are then identified by applying a threshold \( \tau \in [0, 1] \) (e.g., \( \tau = 0.75 \)), producing the set:
\begin{equation}
\mathcal{S}_m = \left\{ i \in \{1, \ldots, L\} \mid \tilde{\alpha}_{m,i} \geq \tau \right\}.
\end{equation}
The lead-level IGAR is defined as:
\begin{equation}
\text{IGAR}_m = \frac{|\mathcal{S}_m|}{L} \times 100,
\end{equation}
where \( L \) is the signal length.

For wave-level, we compute IGAR for each waveform \( w \in \{\text{P}, \text{Q}, \text{R}, \text{S}, \text{T}\} \) detected using NeuroKit2~\cite{makowski2021neurokit2} in each ECG lead \( m \). We calculate the proportion of salient points within the region of waveform \( w \) relative to all salient points in lead \( m \). The IGAR for waveform \( w \) in lead \( m \) is defined as:
\begin{equation}
\text{IGAR}_{m,w} = \frac{|\mathcal{S}_{m,w}|}{|\mathcal{S}_m|} \times 100,
\end{equation}
where \( \mathcal{S}_{m,w} \) is the set of salient points within waveform \( w \) in lead \( m \), and \( \mathcal{S}_m \) is the total set of salient points in lead \( m \). This two-level interpretability approach enables structured analysis of model outputs across leads and clinically meaningful waveform components.

\section{Experiments and Analysis}
In this section, we present our experimental framework to evaluate the capability of our LS-EMVAE model in three binary classification tasks, e.g., PH detection: non-PH (mPAP $\leq$ 20) and PH (mPAP $>$ 20), phenotyping PH: pre-capillary PH (mPAP $>$ 20, PAWP $\leq$ 15) and post-capillary PH (mPAP $>$ 20, PAWP $>$ 15), PAWP prediction: normal (PAWP $\leq$ 15) and elevated (PAWP $>$ 15). We described the datasets, implementation details, and evaluation strategy, followed by comparative experiments that include comparisons between methods and analyses of different ECG lead configurations. We also include a series of ablation studies examining our model’s modules and evaluating its generalizability across different class distributions and genders. We then present both quantitative and qualitative interpretability analyses to support better clinical understanding and transparency of our model’s decision-making in PH detection and phenotyping. Finally, we compare the model complexity of our model with the pre-training baselines and further discuss the broader impact and limitations of our study.

\subsection{Datasets}
\label{sec:dataset}
\subsubsection{Pre-Training Dataset}
We pre-trained our LS-EMVAE model using the \textbf{MIMIC-IV-ECG} dataset~\cite{gowmimic}, which comprises $800,000$ 12L-ECGs from nearly $160,000$ unique patients. The length of the ECGs is 10 seconds with a sampling rate of 500 Hz.

\subsubsection{Fine-Tuning Datasets}
For fine-tuning, we use two retrospective cohorts in a 6L-ECG setting: the \textbf{ASPIRE registry}~\cite{hurdman2012aspire} and \textbf{UK Biobank}~\cite{petersen2016uk}.
\begin{table}[!t]
\caption{Patient characteristics in our ASPIRE registry dataset~\cite{hurdman2012aspire} .}
\label{aspire}
\centering
\resizebox{\linewidth}{!}{
\begin{tabular}{lcccc}
\toprule
\toprule
 & \shortstack{Non-PH\\(mPAP \(\leq 20\))} 
 & \shortstack{PH\\(mPAP \(> 20\))} 
 & \shortstack{Pre-capillary PH\\(PAWP \(\leq 15\))} 
 & \shortstack{Post-capillary PH\\(PAWP \(> 15\))} \\
\midrule
Number of patients & $201$ & $691$ & $136$ & $555$ \\
Age (years) & $58.43\pm11.64$ & $65.41\pm13.02$ & $61.56\pm14.7$ & $69.26\pm11.34$ \\
Male (sex) & $99$ & $177$ & $52$ & $125$ \\
Female (sex) & $102$ & $514$ & $84$ & $430$ \\
Body Surface Area & $1.89\pm0.21$ & $1.93\pm0.26$ & $1.89\pm0.24$ & $1.97\pm0.28$ \\
Heart Rate (bpm) & $68.86\pm8.42$ & $76.94\pm13.99$ & $79.34\pm14.14$ & $74.55\pm13.84$ \\
mPAP (mmHg) & $14.23\pm4.31$ & $26.82\pm5.14$ & $25.32\pm3.86$ & $28.33\pm6.43$ \\
PAWP (mmHg) & $13.67\pm6.77$ & $16.38\pm6.81$ & $11.23\pm2.98$ & $19.53\pm3.85$ \\
\bottomrule
\bottomrule
\end{tabular}
}
\end{table}

\textbf{Study Population of ASPIRE Registry:}
For our primary downstream tasks, PH detection and phenotyping, we evaluate all models using our in-house ASPIRE registry dataset~\cite{hurdman2012aspire}, collected at the Sheffield Pulmonary Vascular Unit. The study was approved by the local institutional review board and ethics committee (16/YH/0352, subsequently 22/EE/0011). A total of 892 patients who underwent RHC and 12L-ECG within 24 hours were included, and only the six limb leads were used in this study. Among these patients, mPAP and PAWP were measured via RHC using a balloon-tipped 7.5-French thermodilution catheter. Based on mPAP, $201$ individuals were identified as healthy and $691$ as having PH. Among these PH patients, $555$ were further classified as pre-capillary PH and $136$ as post-capillary PH based on PAWP. Table~\ref{aspire} provides an overview of the patient characteristics from our ASPIRE registry dataset.

\begin{table}[!t]
\centering
\caption{Hyperparameters in LS-EMVAE pre-training and fine-tuning. The optimal hyperparameter values are shown in \textbf{bold}. $\lambda_{m}$ provides the optimal values for each lead $m$, selected through grid search over \{1, 5, 10\}.}
\label{tab:hyper}
\resizebox{\columnwidth}{!}{%
\begin{tabular}{lrr}
\toprule
Name & Pre-training & Fine-tuning \\ \midrule
Latent dimension & 128/\textbf{256}/512 & 256 \\
Batch size & 64/\textbf{128}/256 & \textbf{32}/64/128\\
$\lambda_m$ & \textbf{(5,10,1,5,1,1,1,10,5,1,1,5)} & $-$ \\
\(\gamma\) & 0.05/\textbf{0.1}/0.5 & $-$\\
Optimizer & AdamW & AdamW\\
Epoch & \textbf{100}/200 & 10/\textbf{50}/100\\
Initial learning rate & 1e-4 & 1e-4\\
Weight decay & 1e-4 & 1e-5 \\
Loss & $\mathcal{L}_{\text{total}}$ & FocalLoss/\textbf{CrossEntropyLoss}  \\ 
FC layer size & $ - $ & 64/\textbf{128}/256  \\ 
Dropout (FC layer) & $ - $ & 0.1/0.3/\textbf{0.5} \\ 
\midrule
GPU &  \multicolumn{2}{r}{NVIDIA A100 NVLink 80GB GPU} \\
\bottomrule
\end{tabular}%
}
\end{table}

\textbf{Study Population of UK Biobank:}
We further fine-tune all models using a larger UK Biobank cohort. In this dataset, cardiac hemodynamics were measured using non-invasive CMRI. However, measurements required to compute mPAP are missing, making direct PH detection unfeasible. Instead, we predict only elevated PAWP, which can indicate post-capillary PH, diastolic dysfunction, or left heart disease. Left atrial volume (LAV) and left ventricular mass (LVM) measured by CMRI were used to calculate PAWP based on the regression formula in~\cite{garg2022cardiac}:
\begin{equation}
\resizebox{0.89\linewidth}{!}{$
\text{CMRI PAWP} = 6.1352 + (0.07204 \times \text{LAV}) + (0.02256 \times \text{LVM}).
$}
\label{eq:cmr-pawp}
\end{equation}
The UK Biobank study was approved by the UK Biobank Research Ethics Committee (11/NW/0382), and our use of the data was approved under application number 99895. A total of 16,416 participants who underwent 12L-ECG before CMRI during the same imaging visit were included in this dataset.

\subsubsection{Dataset Pre-processing}
For both pre-training and fine-tuning datasets, we applied three standard signal preprocessing methods from NeuroKit2~\cite{makowski2021neurokit2} to each ECG lead separately. First, we used the \texttt{signal\_interpolate} method to handle missing or NaN values through linear interpolation. Second, the \texttt{ecg\_clean} method was applied to remove baseline wander and high-frequency noise using a bandpass filter with a frequency range of $0.5$–$40$ Hz. Finally, we used the \texttt{standardize} method to scale each signal to zero mean and unit variance using z-score standardization, ensuring consistent amplitude ranges.
\begin{table*}[!t]
\caption{Performance comparison across three downstream tasks in the 6L-ECG setting, evaluated over 5-fold cross-validation: PH detection and phenotyping (ASPIRE registry) and elevated PAWP prediction (UK Biobank). LS-EMVAE outperforms unimodal methods (multi-lead ECG as a single modality) and multimodal methods (each lead as a separate modality), achieving the best overall results. Results are reported in Accuracy, AUROC, and MCC, with \textbf{best} in bold and \underline{second-best} underlined.}
\label{tab:6l-results}
\centering
\setlength{\tabcolsep}{3pt}
\resizebox{\textwidth}{!}{
\begin{tabular}{ll>{\centering\arraybackslash}p{1.2cm}ccccccccc}
\toprule
\toprule
\textbf{Study Type} & \textbf{Method} & \textbf{Pre-train} 
& \multicolumn{3}{c}{\textbf{Task 1: PH Detection}} 
& \multicolumn{3}{c}{\textbf{Task 2: Phenotyping pre-/post-capillary PH}} 
& \multicolumn{3}{c}{\textbf{Task 3: Elevated PAWP}} \\
\cmidrule(lr){4-6} \cmidrule(lr){7-9} \cmidrule(lr){10-12}
& & 
& \textbf{Accuracy$\uparrow$} & \textbf{AUROC$\uparrow$} & \textbf{MCC$\uparrow$} 
& \textbf{Accuracy$\uparrow$} & \textbf{AUROC$\uparrow$} & \textbf{MCC$\uparrow$} 
& \textbf{Accuracy$\uparrow$} & \textbf{AUROC$\uparrow$} & \textbf{MCC$\uparrow$} \\
\midrule
\multirow{7}{*}{Unimodal}
& Resnet1d~\cite{hong2020holmes} & \xmark 
& $0.733\pm0.09$ & $0.762\pm0.05$ & $0.235\pm0.09$ 
& $0.570\pm0.13$ & $0.763\pm0.05$ & $0.229\pm0.07$ 
& $0.633\pm0.01$ & $0.655\pm0.01$ & $0.206\pm0.01$ \\
& xResnet1d~\cite{strodthoff2020deep} & \xmark 
& $0.772\pm0.01$ & $0.796\pm0.03$ & $0.188\pm0.08$ 
& $0.813\pm0.01$ & $0.798\pm0.06$ & $0.276\pm0.16$ 
& $0.510\pm0.04$ & $0.706\pm0.01$ & $0.218\pm0.04$ \\
& Vanilla Transformer~\cite{vaswani2017attention} & \xmark 
& $0.787\pm0.01$ & $0.771\pm0.02$ & $0.267\pm0.05$ 
& $0.824\pm0.01$ & $0.744\pm0.03$ & $0.352\pm0.06$ 
& $0.661\pm0.01$ & $0.701\pm0.01$ & $0.270\pm0.02$ \\
& $\beta$-VAE~\cite{van2023joint} & \checkmark 
& $0.792\pm0.02$ & $0.795\pm0.02$ & \underline{$0.311\pm0.12$}
& $0.836\pm0.02$ & $0.737\pm0.05$ & $0.394\pm0.10$ 
& $0.635\pm0.01$ & $0.658\pm0.01$ & $0.200\pm0.03$ \\
& ECG-FM~\cite{mckeen2024ecg} & \checkmark 
& \underline{$0.803\pm0.02$} & $0.801\pm0.02$ & $0.283\pm0.04$ 
& $0.848\pm0.01$ & \underline{$0.844\pm0.02$} & $0.428\pm0.04$ 
& $0.665\pm0.01$ & $0.724\pm0.01$ & $0.296\pm0.02$ \\

& ST-MEM~\cite{na2024guiding} & \checkmark 
& $0.791\pm0.01$ & $0.793\pm0.01$ & $0.253\pm0.03$ 
& $0.844\pm0.02$ & $0.816\pm0.01$ & $0.346\pm0.04$ 
& $0.669\pm0.02$ & $0.714\pm0.02$ & $0.286\pm0.03$ \\
& TimesFM~\cite{das2024decoder} & \checkmark 
& $0.802\pm0.01$ & $0.803\pm0.02$ & $0.295\pm0.04$
& $0.851\pm0.02$ & $0.841\pm0.02$ & $0.437\pm0.05$
& $0.667\pm0.02$ & \underline{$0.725\pm0.02$} & $0.306\pm0.04$ \\
\midrule
\multirow{4}{*}{Multimodal}
& MVAE~\cite{wu2018multimodal} & \checkmark 
& $0.790\pm0.02$ & \underline{$0.805\pm0.02$} & $0.274\pm0.12$ 
& \underline{$0.855\pm0.01$} & $0.836\pm0.02$ & \underline{$0.469\pm0.03$} 
& $0.632\pm0.01$ & $0.670\pm0.01$ & $0.164\pm0.08$ \\
& MMVAE+~\cite{palumbo2023mmvae+} & \checkmark 
& $0.746\pm0.01$ & $0.717\pm0.03$ & $0.167\pm0.04$
& $0.814\pm0.01$ & $0.741\pm0.03$ & $0.295\pm0.04$
& \underline{$0.671\pm0.02$} & $0.723\pm0.01$ & \underline{$0.307\pm0.01$} \\
& MoPoE-VAE~\cite{sutter2021generalized} & \checkmark 
& $0.771\pm0.01$ & $0.787\pm0.03$ & $0.258\pm0.10$ 
& $0.836\pm0.01$ & $0.792\pm0.05$ & $0.373\pm0.03$ 
& $0.642\pm0.03$ & $0.709\pm0.02$ & $0.187\pm0.15$ \\
& LS-EMVAE (ours) & \checkmark 
& $\mathbf{0.815\pm0.01}$ & $\mathbf{0.814\pm0.02}$ & $\mathbf{0.412\pm0.05}$ 
& $\mathbf{0.869\pm0.01}$ & $\mathbf{0.863\pm0.03}$ & $\mathbf{0.544\pm0.07}$ 
& $\mathbf{0.682\pm0.01}$ & $\mathbf{0.735\pm0.01}$ & $\mathbf{0.325\pm0.01}$ \\
\bottomrule
\bottomrule
\end{tabular}
}
\end{table*}
\subsection{Implementation and Hyperparameter Setup}
We conduct our experiments on an NVIDIA A100 NVLink 80GB GPU with $256$GB RAM. All models are implemented in Python (version $3.11$) using PyTorch (version $2.3$)~\cite{paszke2019pytorch}. For pre-training, we partitioned the pre-training dataset with a 90:10 split for training and validation, and tuned the hyperparameters using grid search. The optimal configuration was then used to pre-train the model on the full dataset. For fine-tuning, we used 5-fold cross-validation with an 80:20 training and validation split on the fine-tuning datasets. The hyperparameters used in both pre-training and fine-tuning are summarized in Table~\ref{tab:hyper}. 

\subsection{Compared Methods and Evaluation Metrics}
We compared our model with seven established unimodal baselines, including three without pre-training~\cite{strodthoff2020deep, hong2020holmes, vaswani2017attention} and four with pre-training~\cite{van2023joint, mckeen2024ecg, na2024guiding, das2024decoder}. Among these, ResNet1D~\cite{strodthoff2020deep}, XResNet1D~\cite{hong2020holmes}, and $\beta$-VAE~\cite{van2023joint} are CNN-based models that employ input-level ECG lead fusion, while Vanilla Transformer~\cite{vaswani2017attention}, ECG-FM~\cite{mckeen2024ecg}, ST-MEM~\cite{na2024guiding}, and TimesFM~\cite{das2024decoder} are Transformer-based models that perform encoder-level ECG lead fusion using self-attention. XResNet1D and the vanilla Transformer are widely used supervised baselines for ECG classification, whereas ST-MEM, ECG-FM, and Times-FM represent recent strong baselines for ECG and general time-series tasks. We also compared our model with three multimodal latent fusion baselines: MVAE~\cite{wu2018multimodal}, MMVAE+~\cite{palumbo2023mmvae+}, and MoPoE-VAE~\cite{sutter2021generalized}. We benchmarked all the pretrained models using the same 12L-ECG pre-training and 6L-ECG fine-tuning setup as our model. We use three evaluation metrics: accuracy, area under the receiver operating characteristic curve (AUROC), and Matthews correlation coefficient (MCC) across all experiments.

\begin{figure}[t]
\centering
\includegraphics[page=1, width=0.49\textwidth]{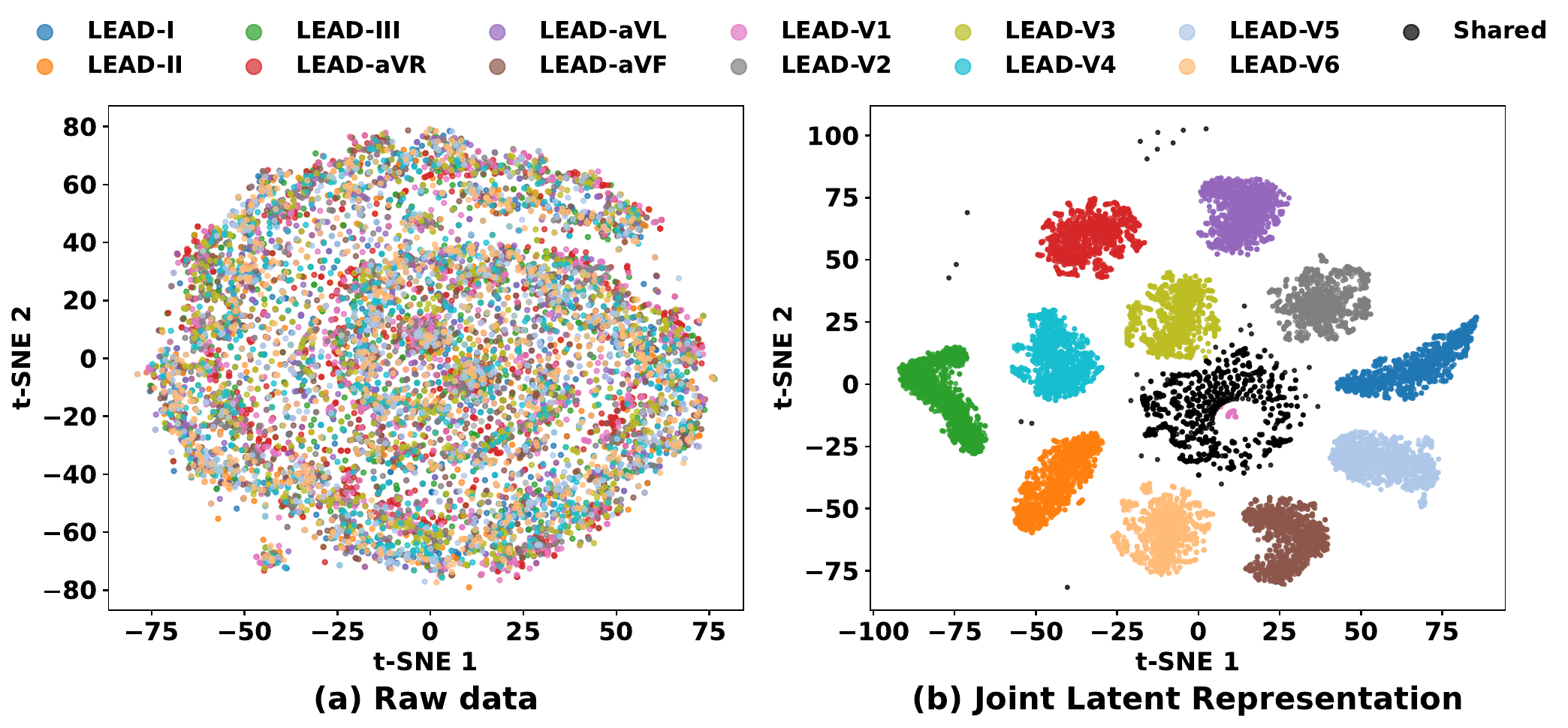}
\caption{T-SNE~\cite{wattenberg2016use} visualization of (a) raw data representation before pre-training and (b) joint representation after pre-training, showing well-formed clusters of lead-specific and shared representations, highlighting effective representation learning by our model.
}
\label{latent}
\end{figure}

\subsection{Comparative Study}

\subsubsection{Performance Comparison with Unimodal Methods}
Table~\ref{tab:6l-results} reports the performance of seven unimodal baselines, where ECG-FM and Times-FM mostly achieve the best performance across tasks. However, the Vanilla Transformer, despite not using any pre-training, still shows reasonably strong performance. Notably, it is also evident that our LS-EMVAE model using multimodal latent fusion consistently outperforms unimodal baselines, which rely on input-level or encoder-level lead fusion. The results show that LS-EMVAE achieves improvements of $\Delta$Accuracy = $0.013$, $\Delta$AUROC = $0.011$, and $\Delta$MCC = $0.117$ for PH detection, $\Delta$Accuracy = $0.018$, $\Delta$AUROC = $0.022$, and $\Delta$MCC = $0.107$ for phenotyping PH, and $\Delta$Accuracy = $0.013$, $\Delta$AUROC = $0.010$, and $\Delta$MCC = $0.019$ for elevated PAWP prediction, compared to the best-performing unimodal baselines. Furthermore, other multimodal models~\cite{wu2018multimodal,palumbo2023mmvae+,sutter2021generalized} that use latent fusion also outperformed unimodal baselines in most cases (5 out of 9). These results confirm that the multimodal latent fusion strategy for ECG leads improves predictive performance compared to unimodal methods.

\subsubsection{Performance Comparison with Multimodal Methods}
As shown in Table~\ref{tab:6l-results}, compared to other multimodal latent fusion models, our model achieves competitive performance with improvements of $\Delta$Accuracy = $0.025$, $\Delta$AUROC = $0.009$, and $\Delta$MCC = $0.138$ for PH detection, $\Delta$Accuracy = $0.014$, $\Delta$AUROC = $0.027$, and $\Delta$MCC = $0.075$ for phenotyping PH, and $\Delta$Accuracy = $0.011$, $\Delta$AUROC = $0.012$, and $\Delta$MCC = $0.018$ for elevated PAWP prediction. Among these models, only LS-EMVAE effectively handles both individual lead and shared representations, resulting in a well-clustered joint latent representation as shown in Fig.~\ref{latent}. Its superior performance indicates that handling these representations effectively during pre-training leads to better generalization during fine-tuning. 

\begin{figure}[t]
\centering
\includegraphics[page=1, width=0.49\textwidth]{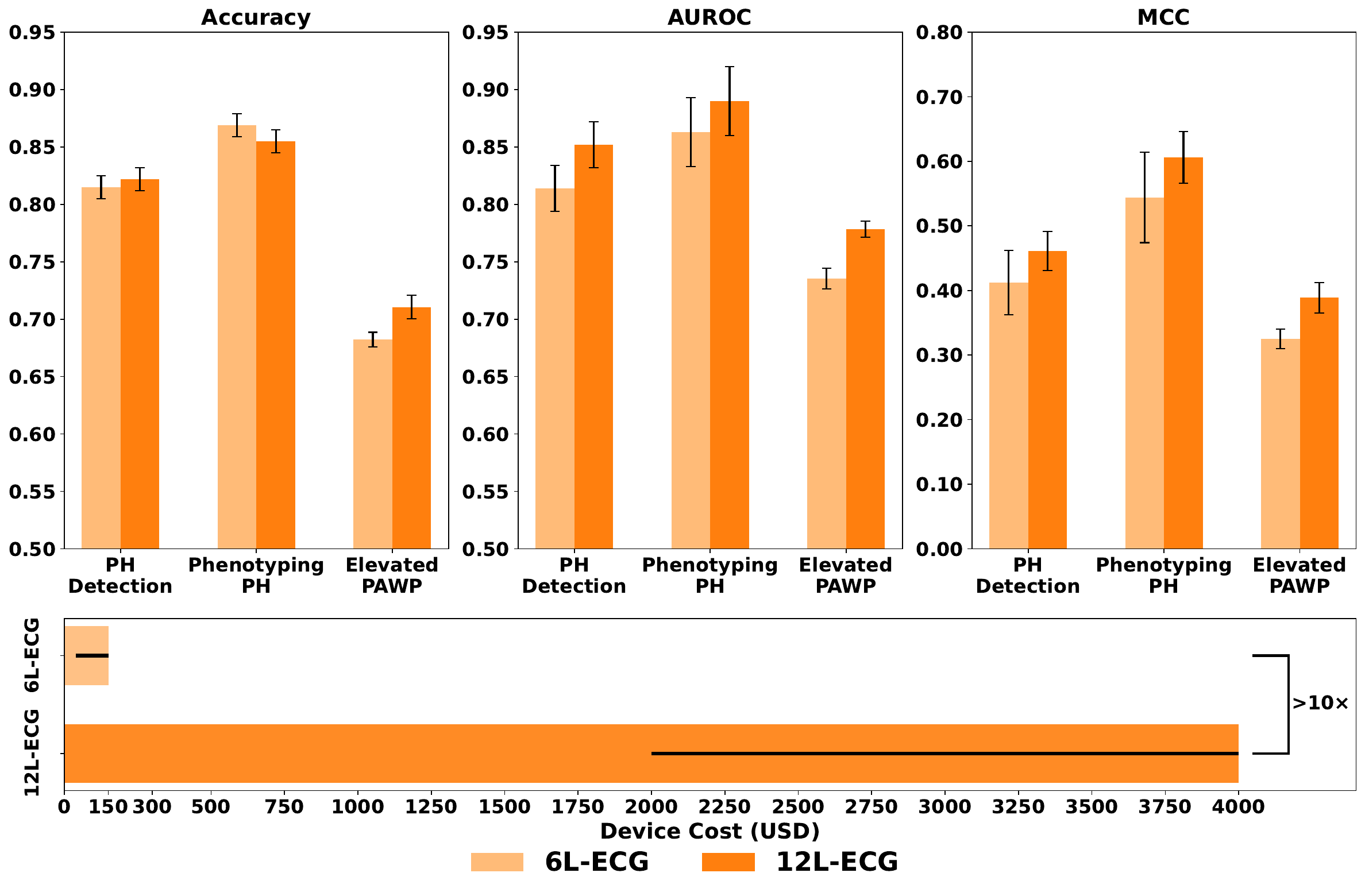}
\caption{Performance of LS-EMVAE across 6L-ECG and 12L-ECG settings, where 12L-ECG achieves higher performance, while 6L-ECG offers a cost-effective alternative, underscoring the trade-off between performance and cost.}
\label{6v12}
\end{figure}

\begin{figure*}[t]
\centering
\includegraphics[page=1, width=\textwidth]{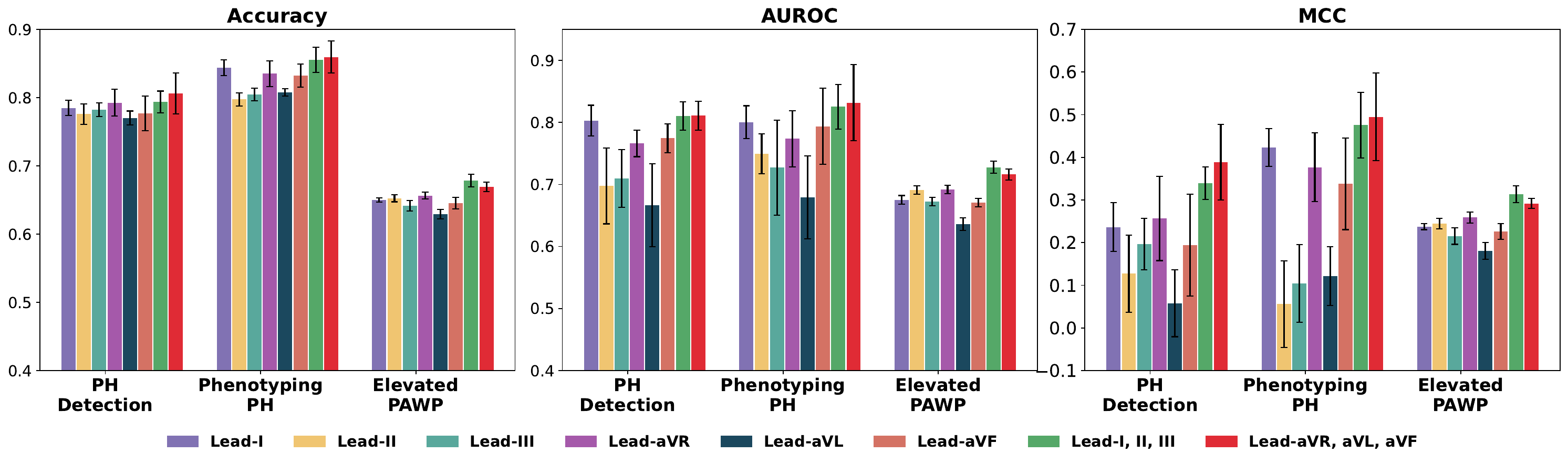}
\caption{Performance of LS-EMVAE on reduced ECG configurations comprising individual leads, bipolar limb leads (I, II, III), and augmented limb leads (aVR, aVL, aVF). Lead-I and Lead-aVR achieve the strongest performance among individual leads, while augmented leads outperform bipolar leads in most cases (6 out of 9), highlighting their clinical relevance.}
\label{ablation:lead-analysis}
\end{figure*}

\subsubsection{Performance Comparison between 6L-ECG and 12L-ECG Settings}
We further analyze our model’s performance using the 12L-ECG setting compared to the 6L-ECG setting, as shown in Fig.~\ref{6v12}, where in most cases 12L-ECG achieved better performance. Specifically, the observed performance differences are $\Delta$Accuracy = $0.007$, $\Delta$AUROC = $0.038$, and $\Delta$MCC = $0.049$ for PH detection, $\Delta$Accuracy = $0.014$, $\Delta$AUROC = $0.021$, and $\Delta$MCC = $0.062$ for PH phenotyping, and $\Delta$Accuracy = $0.028$, $\Delta$AUROC = $0.043$, and $\Delta$MCC = $0.063$ for elevated PAWP prediction. However, a 6L-ECG device costs less than $150$, whereas a 12L-ECG device costs between $2,000$ and $4,000$, which is over ten times more expensive~\cite{slater2025validation}, highlighting a trade-off between performance and cost. This balance supports adopting 6L-ECG with our LS-EMVAE model as a practical and scalable solution for PH assessment in decentralized clinical settings.

\subsubsection{Comparison of different ECG Lead Configurations}
\label{section:ecg-analysis}
Fig.~\ref{ablation:lead-analysis} shows the performance of our model using individual lead configurations from the 6L-ECG. We also evaluate standard ECG configurations, including the bipolar limb leads (I, II, III) and the augmented limb leads (aVR, aVL, aVF). Among the individual leads, Lead-I and Lead-aVR show the strongest performance. This is particularly promising for Lead-I, as it captures signals from the arms and is widely used in wearable ECG devices, making it suitable for continuous cardiac monitoring. This finding motivates future research into the use of ECG devices relying solely on Lead-I for detecting and phenotyping PH in continuous monitoring settings. Lead-aVR is also noteworthy, as it performs strongly despite often being considered the “forgotten lead” in clinical practice~\cite{george2010avr}. It is an augmented limb lead theoretically derived from Lead-I and Lead-II, represented as $-1/2 \times (\text{Lead-I} + \text{Lead-II})$. To further assess the role of augmented leads, we compared them with bipolar limb leads. For PH detection and phenotyping, augmented leads outperform bipolar leads in most cases (6 out of 9), underscoring their clinical relevance.

\subsection{Ablation Study}
\subsubsection{Effects of Different Components in LS-EMVAE}
We conduct an ablation study across the three downstream tasks by individually removing PoE, MoE, the LRA loss, and pre-training to evaluate their contributions to model performance. In Table~\ref{tab:ablation-auroc}, the results show that excluding PoE from LS-EMVAE significantly reduces performance ($\Delta$AUROC = $0.082$, $0.143$, and $0.127$), as the model can no longer effectively incorporate lead shared information into the joint latent representation. When MoE is excluded, performance also drops substantially ($\Delta$AUROC = $0.078$, $0.148$, and $0.153$), likely because the model fails to effectively capture individual lead information. Further, the removal of the LRA loss causes a notable drop ($\Delta$AUROC = $0.018$, $0.052$, and $0.038$), as this weakens the alignment between individual leads and shared representation. Finally, without pre-training, the model shows the lowest performance, with AUROC reduced by $0.158$, $0.309$, and $0.049$. These results demonstrate the essential role of all three modules and the pre-training stage in enhancing the model’s generalization and diagnostic performance.

\begin{table}[t]
\centering
\caption{Effects of PoE, MoE, LRA, and pre-training in the LS-EMVAE. Each value indicates the mean AUROC score with standard deviation.}
\label{tab:ablation-auroc}
\scalebox{0.82}{
\begin{tabular}{>{\centering\arraybackslash}p{0.4cm} >{\centering\arraybackslash}p{0.4cm} >{\centering\arraybackslash}p{0.4cm} >{\centering\arraybackslash}p{1.2cm}>{\centering\arraybackslash}p{1.6cm}>{\centering\arraybackslash}p{1.6cm}>{\centering\arraybackslash}p{1.6cm}}
\toprule\toprule
\textbf{PoE} & \textbf{MoE} & \textbf{LRA} & \textbf{Pre-train} & \textbf{PH Detection$\uparrow$} & \textbf{Phenotyping PH$\uparrow$} & \textbf{Elevated PAWP$\uparrow$} \\
\midrule
\xmark      & \checkmark   & \checkmark   & \checkmark      & $0.732\pm0.04$       & $0.720\pm0.02$ & $0.608\pm0.01$ \\
\checkmark  & \xmark       & \checkmark   & \checkmark      & $0.736\pm0.03$                  & $0.715\pm0.04$ & $0.582\pm0.01$\\
\checkmark  & \checkmark   & \xmark       & \checkmark      & $0.796\pm0.03$       & $0.811\pm0.04$ & $0.687\pm0.01$\\
\checkmark  & \checkmark   & \checkmark   & \xmark          & $0.656\pm0.04$       & $0.554\pm0.02$ & $0.686\pm0.01$\\
\checkmark  & \checkmark   & \checkmark   & \checkmark      & $\mathbf{0.814\pm0.02}$ & $\mathbf{0.863\pm0.03}$ & $\mathbf{0.735\pm0.01}$\\
\bottomrule\bottomrule
\end{tabular}
}
\end{table}

\begin{figure}[t]
\centering
\includegraphics[page=1, width=0.49\textwidth]{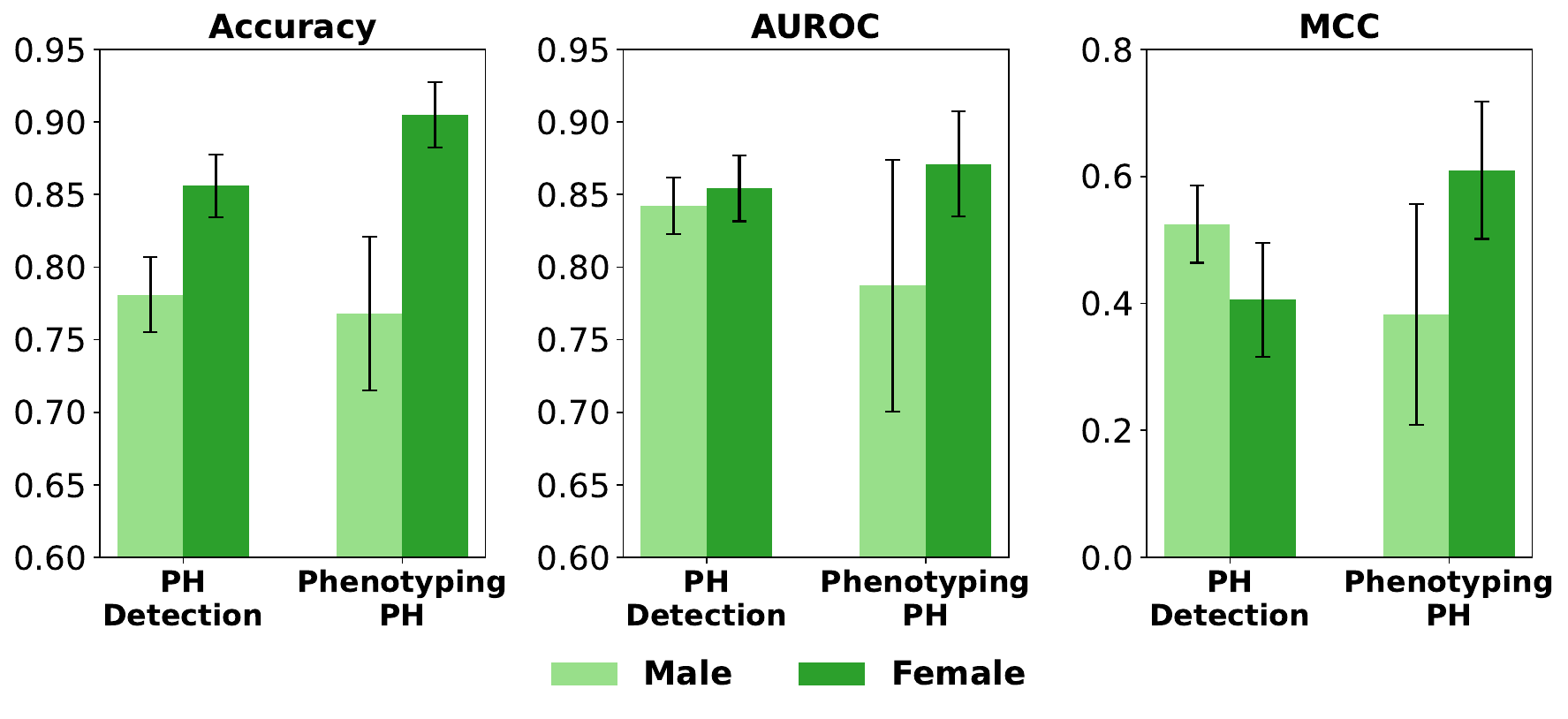}
\caption{Performance of LS-EMVAE on PH detection and phenotyping across genders, showing reasonable performance in both groups with variations that suggest some gender-specific differences.}
\label{abalation:gender}
\end{figure}

\begin{figure}[t]
\centering
\includegraphics[page=1, width=0.49\textwidth]{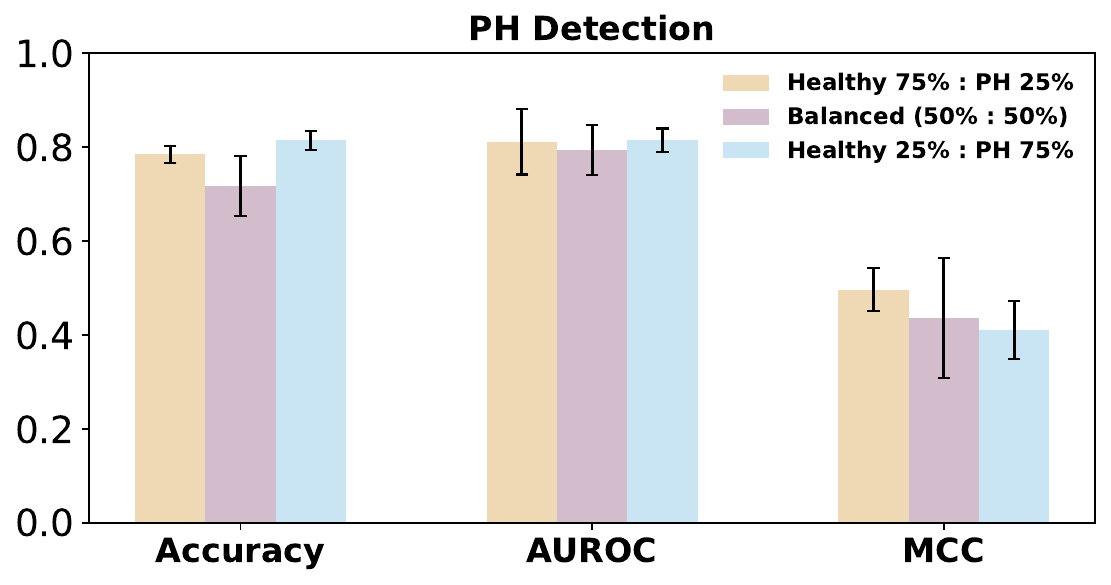}
\caption{Performance of LS-EMVAE across different class distributions for PH detection, showing consistent results under real-world, balanced, and PH-dominant settings.}
\label{abalation:class}
\end{figure}

\subsubsection{Generalizability of LS-EMVAE across Genders}
Fig.~\ref{abalation:gender} presents the performance of our model on male and female patients from the ASPIRE registry for PH detection and phenotyping. The model consistently performs better on female patients, likely due to the gender imbalance in the dataset, where female samples outnumber male samples, as shown in Table~\ref{aspire}. Clinically, PH is more prevalent in females~\cite{rodriguez2021sex}, which explains the higher proportion of female samples in the dataset. Nevertheless, our model maintains reasonable performance across both genders, although with notable variations, suggesting a degree of generalization but some gender-specific differences.

\subsubsection{Generalizability of LS-EMVAE Across Different Class Distributions}
Fig.~\ref{abalation:class} shows the performance of our model on PH detection across three different class distribution settings. The first setting reflects a real-world scenario where healthy patients are more prevalent than PH cases. The second uses a balanced distribution of healthy and PH cases, while the third mirrors our dataset, where PH cases are more frequent. Across all three distribution settings, LS-EMVAE demonstrates consistent performance, highlighting its strong generalization capability and robustness to class imbalance and distribution shifts.

\subsection{LS-EMVAE Model Interpretation for PH Detection and Phenotyping}
In this subsection, we present the interpretation of our LS-EMVAE model for PH detection and phenotyping, performing both quantitative and qualitative interpretations. These interpretations aim to provide clinically meaningful insights and build confidence in the model’s predictions.

\begin{figure}[t]
\centering
\includegraphics[page=1, width=0.49\textwidth]{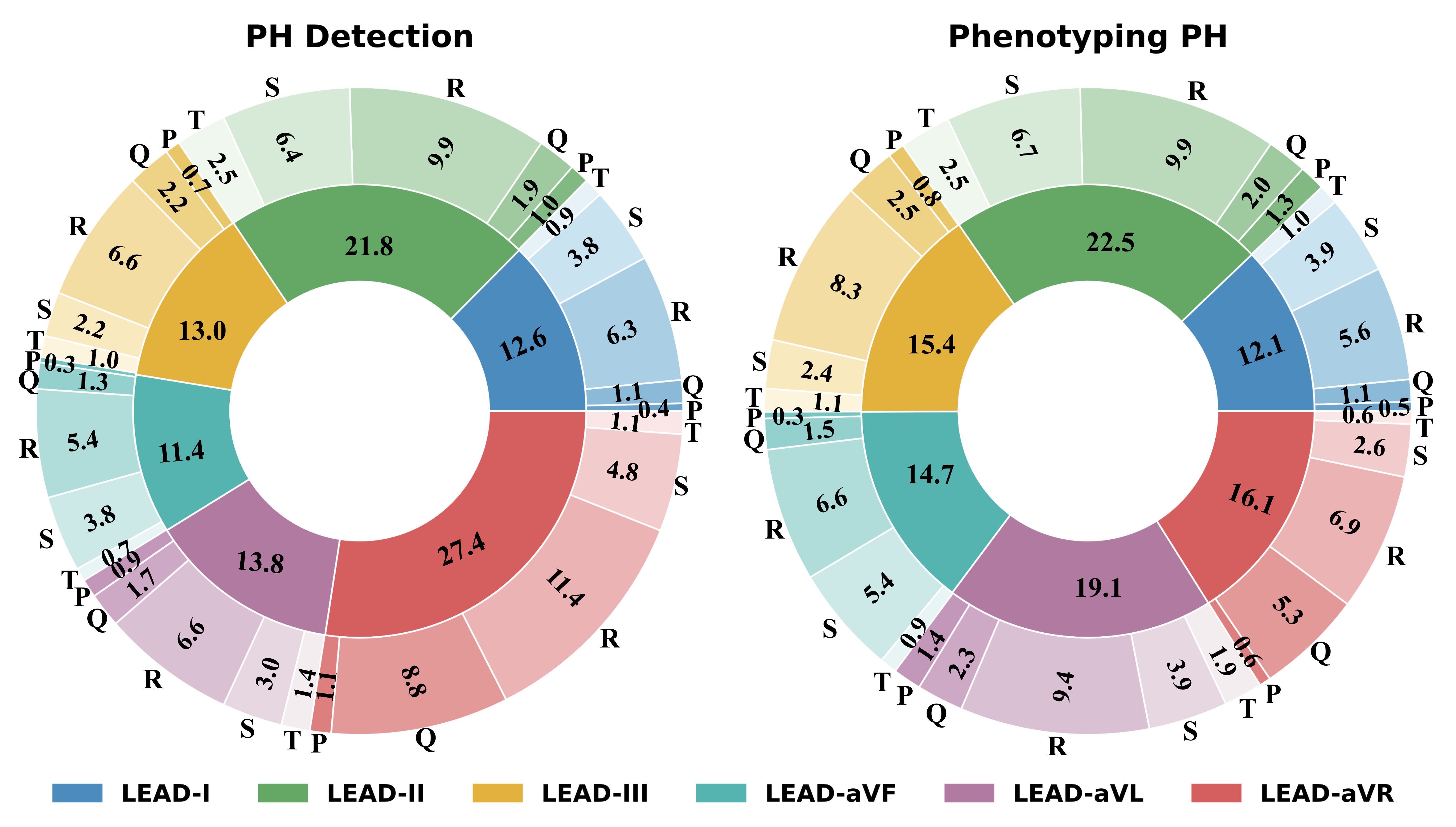}
\caption{Quantitative interpretation of LS-EMVAE using IGAR on PH detection and phenotyping. Each nested pie chart displays the contribution of 6L-ECG leads (inner ring) and ECG waveforms P, Q, R, S, T (outer ring) over 5-fold cross-validation. The interpretation shows that Lead-aVR and Lead-II are the most informative leads, while the QRS complex (Q, R, S waves) contributes the most at the waveform level. All the values are in percentage (\%).}
\label{quant}
\end{figure}

\begin{figure*}[!t]
\centering
\includegraphics[page=1, width=\textwidth]{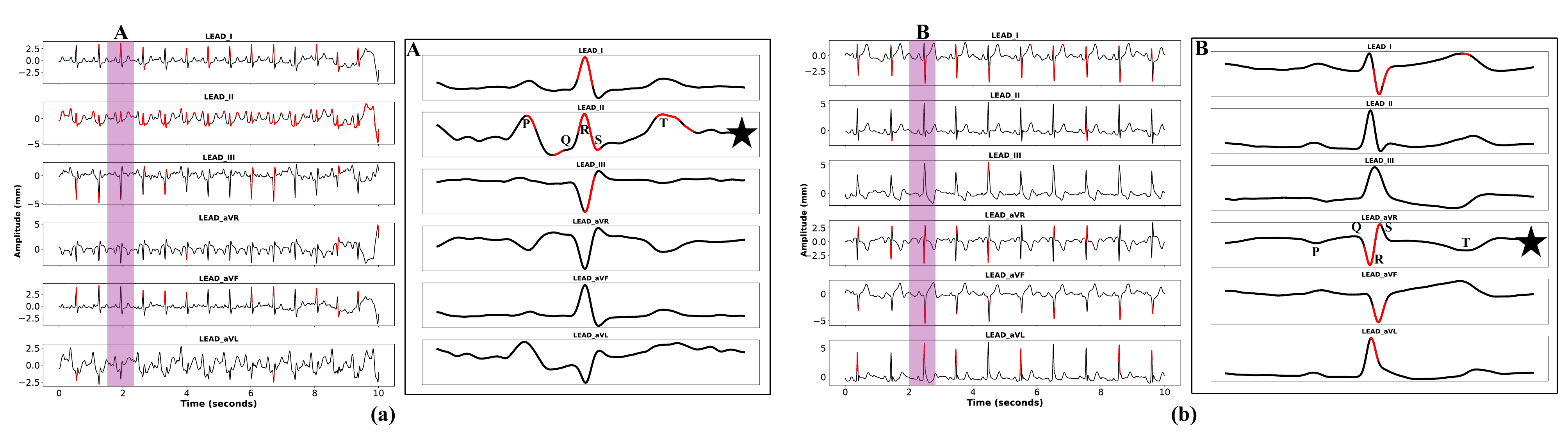}
\caption{Qualitative interpretation of LS-EMVAE for two PH patients (a) and (b) identified by our model, highlighting Lead-II (\(\bigstar\)) and Lead-aVR (\(\bigstar\)) as the most influential leads in PH detection.}
\label{interpret}
\end{figure*}

\subsubsection{Quantitative Analysis}
We used the IGAR method to quantitatively interpret our model, assessing its decisions across the validation set rather than relying on a few cherry-picked examples that could misrepresent overall performance. In Fig.~\ref{quant} (inner circle), we present the mean IGAR results obtained over 5-fold cross-validation for PH detection and phenotyping, showing the contribution of each 6L-ECG lead. In PH detection, Lead-aVR contributes the most, followed by Lead-II, which aligns with the findings in Section~\ref{section:ecg-analysis} and further confirms its strong relevance to PH detection. Clinically, its relevance is linked to its ability to capture right-sided electrical activity, which is commonly affected in PH patients. In phenotyping PH, Lead-II clearly stands out, consistent with its established clinical relevance as the most informative lead\cite{henkens2008pulmonary}. Notably, all six leads contributed more than 10\%, underscoring that each lead provides distinct information and is important for PH detection and phenotyping.

In Fig.~\ref{quant} (outer circle), we further present our findings on the importance of each ECG waveform (P, Q, R, S, and T) across all six leads. Unlike lead-level interpretation, wave-level interpretation offers greater clinical insight, as these waveforms represent key phases of the cardiac cycle. They also reflect specific electrical activities and are primary features clinicians use to detect structural or functional heart changes. In our waveform analysis, the R wave consistently contributes the most, which is expected since it represents ventricular depolarization and usually dominates the ECG signal across various cardiac conditions. In both PH detection and phenotyping tasks, Lead-II shows that the S wave is the second most important, contributing around 6.4\% and 6.7\% respectively. This is clinically relevant because a prominent S wave in Lead-II suggests right ventricular strain or hypertrophy, which are commonly observed in PH. In Lead-aVR, the Q wave is the second most important, contributing 8.8\% in PH detection and 5.3\% in phenotyping. This reflects early septal depolarization changes that become more evident in augmented leads, especially in patients with right-sided heart pressure overload. Overall, we found that the Q, R, and S waves contributed the most across all leads. This highlights the importance of the QRS complex region, which captures the complete process of ventricular depolarization. Since PH often leads to structural and conduction changes in the right ventricle, abnormalities in the QRS complex serve as key indicators for both PH detection and phenotyping. 

\subsubsection{Qualitative Analysis}
In Fig.~\ref{interpret}, we use IG~\cite{sundararajan2017axiomatic} for qualitative interpretation of two PH patients identified by our model. The red-highlighted regions indicate the most influential parts of the 6L-ECG signals for the model’s prediction. In both examples, the QRS complex region dominates across leads. However, without the quantitative IGAR interpretation, we would not know that some leads, although less prominent in these two cases, still make notable contributions in PH assessment. This underscores the importance of combining qualitative and quantitative interpretation to better understand how the model learns and makes decisions, thereby supporting more robust findings and minimizing interpretation bias.

\subsection{Model Complexity}
Table~\ref{tab:model-complexity} compares the complexity of our model against the pre-trained baseline models. We assess model complexity using three metrics: the number of parameters, floating-point operations (FLOPs), and inference time. Since all multimodal models, including ours, treat each ECG lead as a separate modality and use separate encoders, they have a larger number of parameters than unimodal models. However, ECG sequences are relatively short, and each parameter is used only once per sample. As a result, the computational burden remains low, and LS-EMVAE achieves faster inference time. In practice, our model is memory-bound rather than compute-bound. All measurements were obtained on a single NVIDIA A100 80GB GPU, and inference time may vary across hardware.

\begin{table}[t]
\centering
\caption{Model complexity comparison: parameter count (millions), computational cost (GFLOPs), and inference time (ms, batch size = 1). LS-EMVAE has more parameters than unimodal models but maintains low computational cost and faster inference time.}
\label{tab:model-complexity}
\scalebox{0.85}{
\begin{tabular}{lccc}
\toprule\toprule
\textbf{Model} & \textbf{Params ($10^6$)}$\downarrow$ & \textbf{FLOPs ($10^9$)}$\downarrow$ & \textbf{Inference Time (ms)}$\downarrow$ \\
\midrule
$\beta$-VAE~\cite{van2023joint}              & $$25$$     & $$0.50$$    & $$2$$    \\
ECG-FM~\cite{mckeen2024ecg}                  & $$90.9$$   & $$13.00$$   & $$18$$   \\
ST-MEM~\cite{na2024guiding}                  & $$86.0$$   & $$11.00$$   & $$16$$   \\
Times-FM~\cite{das2024decoder}               & $$200.0$$  & $$20.00$$   & $$25$$   \\
MVAE~\cite{wu2018multimodal}                 & $$256.0$$  & $$0.68$$    & $$2$$    \\
MMVAE+~\cite{palumbo2023mmvae+}              & $$257.0$$  & $$0.70$$    & $$2$$    \\
MoPoE-VAE~\cite{sutter2021generalized}       & $$256.0$$  & $$0.69$$    & $$2$$    \\
\midrule
\textbf{LS-EMVAE (ours)}                     & $$256.0$$  & $$0.70$$    & $$2$$    \\
\bottomrule\bottomrule
\end{tabular}
}
\end{table}

\subsection{Broader Impact \& Limitations}
Our study addresses a critical gap in decentralized healthcare by enabling reliable PH detection and phenotyping using 6L-ECG. As a low-cost and portable alternative to 12L-ECG, it is well-suited for GP, POC, and home-based care, where clinical decisions often rely solely on symptom-based evaluation. By supporting early PH assessment in these environments, LS-EMVAE could improve diagnosis rates in low- and middle-income countries (LMICs), where up to 80\% of cases remain undiagnosed~\cite{maarman2020pulmonary}. This would enable earlier intervention and help reduce both the clinical burden and long-term healthcare costs worldwide. While our results demonstrate that 6L-ECG can effectively support PH detection and phenotyping, we have not directly confirmed its performance in handheld 6L-ECG devices. This work should therefore be considered a pre-trial investigation, providing evidence to inform future prospective studies using handheld 6L-ECG data in decentralized settings.

\section{Conclusion}
\label{sec:conclusion}
This paper presents a multimodal latent fusion model for ECG leads in PH assessment. We demonstrated that (1) Our multimodal approach outperforms existing unimodal and multimodal baselines across downstream tasks by learning a rich joint latent representation that integrates individual lead and shared information, effectively improving performance and generalization, (2) using our model, 6L-ECG provides a cost-effective alternative that achieves strong performance in PH detection and phenotyping, making it suitable for deployment in decentralized clinical settings such as point-of-care, community healthcare, and home care, and (3) the interpretability offered by our model aligns with clinical knowledge and supports clinical decision-making, with findings consistent with established clinical practice.


\section*{References}
\vspace{-19pt}
\bibliographystyle{IEEEtran}
\bibliography{LSEMVAE.bib}

\end{document}